\documentclass[sigconf, 10pt, authorversion, nonacm]{acmart}
\usepackage{graphicx} %

\definecolor{brewerpurple}{HTML}{AF4EA3}
\definecolor{brewerblue}{HTML}{377EB8}
\definecolor{NavyBlue}{HTML}{006EB8}
\definecolor{BrickRed}{HTML}{B6321C}
\definecolor{ForestGreen}{HTML}{009B55}
\usepackage{listings}

\lstdefinestyle{customc2}{
    emphstyle=\bfseries\color{NavyBlue},
    commentstyle=\color{ForestGreen}\itshape\ttfamily,
       morekeywords={yield},
    stringstyle=\color{red}\ttfamily,
    emph={[2]admission\_policy, scheduling\_policy, placement\_policy},emphstyle={[2]\bfseries\color{BrickRed}},
    language=Python,
    keywordstyle=\bfseries\color{green!40!black}
}
\author{Saurabh Agarwal}
\affiliation{
\institution{University of Texas-Austin}
\country{}
}

\author{Marco Laju}
\affiliation{
\institution{University of Texas-Austin}
\country{}
}
\author{Jayanth Srinivasa}
\affiliation{
\institution{Cisco-Research}
\country{}
}
\author{Myungjin Lee}
\affiliation{
\institution{Cisco-Research}
\country{}
}
\author{Aditya Akella}
\affiliation{
\institution{University of Texas-Austin}
\country{}
}

\usepackage{mathtools}
\usepackage{multirow}
\usepackage{wrapfig}
\usepackage{url}
\usepackage{xcolor}
\usepackage{enumitem}
\usepackage{capt-of}
\usepackage{booktabs}
\usepackage{subcaption}
\usepackage{calrsfs}

\usepackage{xspace}
\usepackage{soul}
\usepackage[noabbrev]{cleveref}

\newcommand{\eg}{{\it e.g.}, }

\usepackage{graphicx}
\graphicspath{ {./images/} }

\begin{document}
\title{Software-Defined Agentic Serving}

\begin{abstract}
As multi-agent LLM pipelines grow in complexity, existing serving paradigms fail to adapt to the dynamic serving conditions. We argue that agentic serving systems should be programmable and system-aware, unlike existing serving which statically encode the parameters. 
In this work, we propose a new SDN-inspired agentic serving framework that helps control the key attributes of communication based on runtime state. This architecture enables serving-efficient, responsive agent systems and paves the way for high-level intent-driven agentic serving.

\end{abstract}
\maketitle

\section{Introduction}

The rise of Large Language Models (LLMs) has catalyzed a growing ecosystem of intelligent LLM-powered agents—each acting as a specialized model for tasks such as code generation, testing, retrieval, and planning. To accomplish these complex tasks, these  agents need to 
routinely invoke APIs, tools, and other models in "agentic pipelines"~\cite{park2023generative,talebirad2023multi}. 

\begin{figure*}[t]
\begin{minipage}[t]{0.32\textwidth}
   \includegraphics[width=0.9\linewidth]{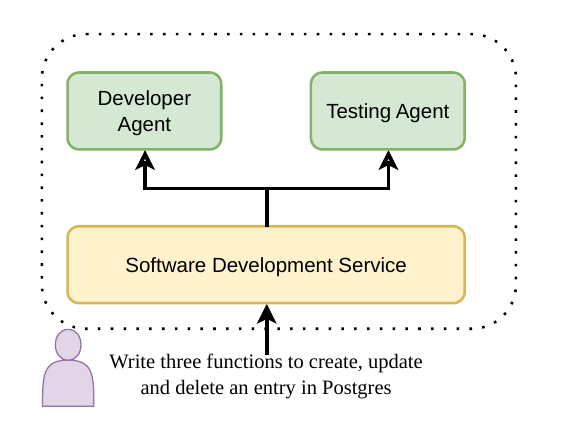}
   \vspace{-10pt}
    \caption{\small\textbf{Agentic Software Developer:} The above highlights a software agent workflow where the developer agent is responsible for generating functions and the testing agent generates testing code. }
    \label{fig:workflow_agent} 
\end{minipage}\quad
\begin{minipage}[t]{0.32\textwidth}
    \includegraphics[width=0.9\linewidth]{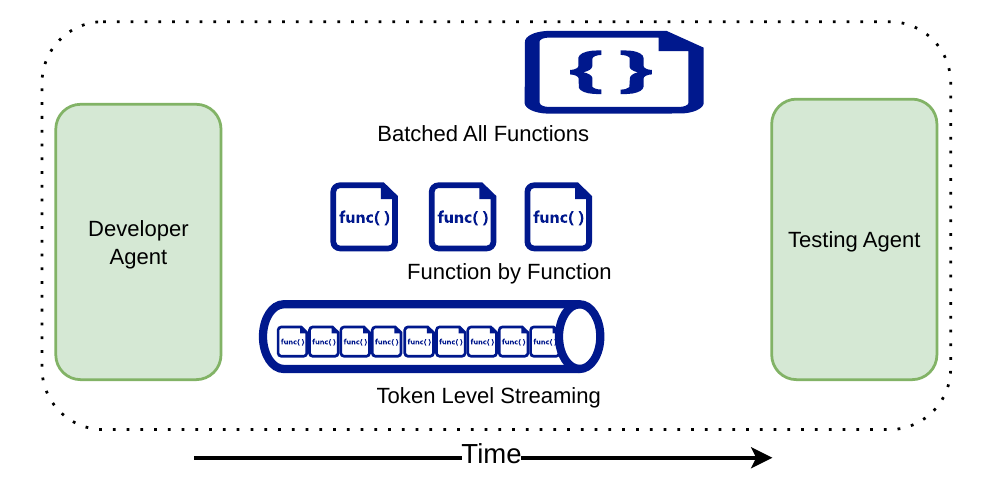}
    \vspace{-10pt}
    \caption{\small\textbf{Communication Pattern:} The above schematic provides an approximate timeline for the communication pattern for a possible different communication pattern.}
    \label{fig:commpattern}
\end{minipage}\quad
\begin{minipage}[t]{0.32\textwidth}
\includegraphics[width=0.9\linewidth]{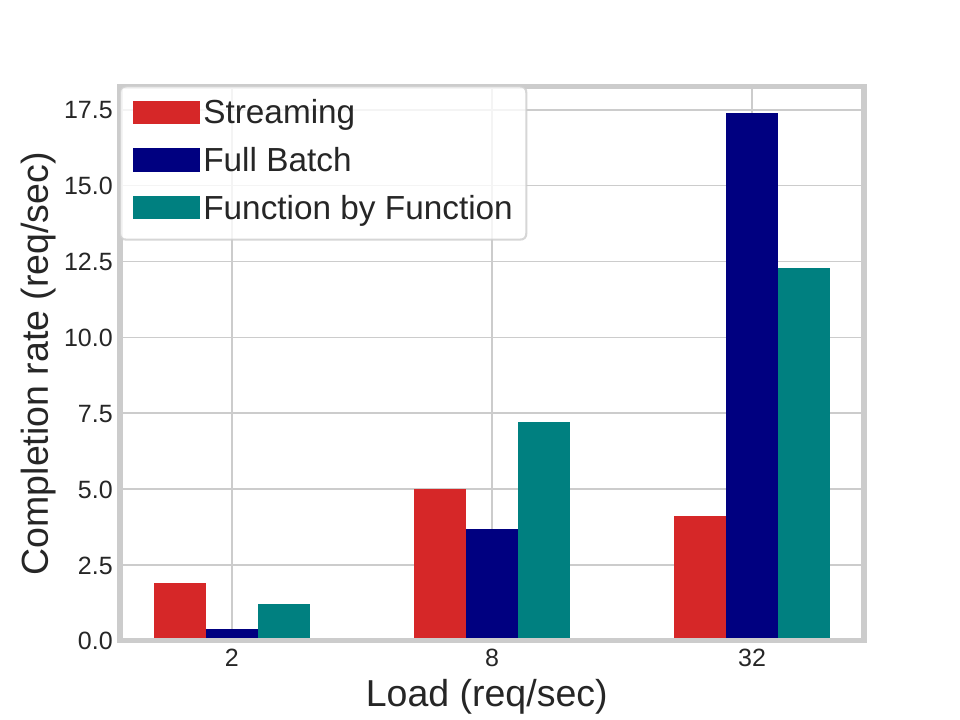}
    \vspace{-13pt}
    \caption{\small{\textbf{Serving throughput using A2A:} We serve two agents using three different communication mechanisms under varying load. No one configuration consistently outperforms another.}} %
    \label{fig:request_rate}
\end{minipage}
\end{figure*}

Consider an agentic software development pipeline illustrated in Figure~\ref{fig:workflow_agent}. The workflow involves communication between two agents: a "developer" agent that emits functions, and a "tester" agent that provides feedback. 
There are several possible communication strategies between the agents: (a) \emph{Batching} all functions in one call, which can reduce overhead and improves throughput. (b) \emph{Function-by-function} pipelining, which enables early feedback and concurrency. (c) \emph{Token-level streaming}, which minimizes latency for interactive responsiveness.
In this example, the “best” strategy depends as much on the application needs as on the {\em current system state}. For example, batching may preserve throughput under high load and reduce backlogs; streaming improves app responsiveness under low load. 
More crucially, the agnetic serving decisions--- regarding batch size, routing, communication granularity---both influence %
and are influenced by runtime system %
factors, e.g., load, queue lengths, model latency, token generation rates, user responsiveness goals, SLOs, etc. (\S\ref{sec:motivation}). Ideally, the decisions should thus be {\em dynamic and adaptive to serving-system state.}

Unfortunately, today's agentic serving architectures make optimal strategy selection difficult, especially as system dynamics change. %
This is due to three fundamental drawbacks in existing agentic serving designs. 
\emph{First}, agentic communication protocols today, such as Anthropic’s Model Context Protocol (MCP)~\cite{mcp}, Google’s Agent2Agent (A2A)~\cite{a2a}, IBM’s Agent Connection Protocol (ACP)~\cite{acp}, and others~\cite{agntcy}, 
require developers to \emph{early bind} to a strategy at design time---such as choosing batching, streaming, or pipelined sends.  
\emph{Second}, serving parameters like batch size, communication frequency, routing strategy, model selection are {\em unaware of serving-layer parameters like} -- model latency, queue times, cost per inference, or cross-agent resource contention --
on which performance critically depends.  
\emph{Finally}, today's communication architectures {\em lack interfaces for cross-agent and pipeline-wide optimization},  e.g., for prioritization at downstream agents, or balancing user experience with end-to-end system throughput. 

Taken together, these attributes make agentic serving architectures both a burden on developers of agentic pipelines--it is difficult to ensure and control performance--and a source of serious run-time inefficiency. %

{\em We advocate for a new agentic serving architecture where the agent communication is dynamically controllable and strongly coupled with the serving system.} Specifically, inspired by SDN, we argue for a programmable agentic substrate. However, distinct from SDN, we require the substrate to go beyond communication and also interface directly with applications to enable better control over serving, ensuring efficiency and performance. %

We envision a "software-defined agentic serving stack" with three components:
\begin{itemize}
    \item A \textbf{data plane} that supports fine-grained, dynamically adjustable agentic message granularities spanning token-level streaming to batched contexts. 
    \item  A \textbf{metrics plane} which provides low-overhead access to metrics via a simple, cross-agent/tool uniform data collection interface. 
    \item A \textbf{control plane} that leverages metrics to dynamically govern inter-agent communication strategies, batch size, routing, and model selection across agent pipelines. It systematically evolves system-wide state to a target that meets operator-specified goals. Control decisions account for both system- and application-level metrics.
\end{itemize}

We envision such a stack to support high-level declarative languages for agent communication and serving policy. These languages could allow infrastructure engineers to express goals---such as "minimize end-to-end latency," "batch unless load is low," or "stream only for interactive or high-priority tasks"---without needing to manage the low-level implementation details. The control plane translates these intents into real-time decisions based on the current serving state. The architecture frees agentic workflow developers from implementing bespoke approaches for controlling application performance; instead, agentic pipelines simply offer the control plane the appropriate hints.

We discuss challenges arising from tool/agent heterogeneity and churn in realizing the above vision. We present simple unified interfaces and design ideas for the three planes that can enable effective control despite these attributes.

Our preliminary prototype shows that our architecture can improve agentic serving throughput by up to $3.6\times$ by fine-grained control of the communication granularity. Further, when additional control is exercised over the serving system, the serving throughput improves by another $2.3\times$.

More than a specific proposal, the main aim of our work is to kickstart a discussion around addressing fundamental issues in emerging agentic architectures. We present a framework inspired by SDN, along with APIs that integrate with current and future tools and agents, to overcome these issues. But we recognize that these are mere starting points in an interesting design space. We conclude by identifying key future directions in \S\ref{sec:discussion}.

\section{Agentic Serving }
\label{sec:motivation}
\begin{figure}
    \centering
   
\end{figure}

LLMs are rapidly evolving beyond 
their basic use as chatbots, increasingly serving as full-scale digital assistants.
In particular, LLMs are being given access to additional tools~\cite{qin2023toolllm, yang2023gpt4tools, schick2023toolformer} (\eg calculators, web search engines, databases, file systems, to name a few) and being treated as autonomous agents cooperating with other agents to fulfill complex tasks (e.g., software development~\cite{hong2023metagpt} and problem solving~\cite{tan2024taskgen}). %
Such workflows necessitate LLMs communicating with other tools and LLMs.

Unfortunately, today's {\em agentic serving systems} have fundamental limitations. We start by describing how these systems work.

\begin{figure}[t!]
\centering
\resizebox{0.95\linewidth}{!}{
\begin{lstlisting}[style=customc2]
# Client initialization and capability determination
with httpx.client() as httpx_client:
    developer_agent = A2AClient.get_client_from_agent_card_url(
    httpx_client, "http://localhost:6069"
)   

    testing_agent = A2AClient.get_client_from_agent_card_url(
    httpx_client, "http://localhost:6070"
)  
    prompt = "Write three functions to creating an entry, 
    update it and delete it"
    
    send_message_payload = message{
        'role': "user",
        'parts': [{'type': 'text', 'text': prompt}],
    }
   
    streaming_request = SendStreamingMessageRequest(
                params=MessageSendParams(**send_message_payload)
            ) 
	# perform streaming message 
    stream_response = developer_agent.send_message_streaming(
        streaming_request)
	# waiting for streaming response
    async for chunk in stream_response:
        # stream to testing agent
         testing_request =SendStreamingMessageRequest(
                params=MessageSendParams(**chunk)
        )
        test_funtion = testing_agent.send_message_streaming(chunk)
        await asyncio.sleep(0.1)
\end{lstlisting}}
\vspace{-10pt}
\caption{\small{\textbf{A code snippet illustrating an agentic application written using A2A.}}}
\label{fig:code_listing}
\vspace{-10pt}
\end{figure}

\subsection{ Agentic Workflows}
\label{sec:limitations}

Constructing multi-agent LLM workflows today requires developers to manually orchestrate communication across tools and models.  Developers curate a list of callable agents, handling authentication, and managing communication modes for each endpoint—often via heterogeneous mechanisms such as REST, RPC, or streaming APIs.
To manage this complexity, recent efforts have introduced protocol-based abstractions. Examples include Google's A2A~\cite{a2a}, Anthropic's MCP~\cite{mcp}, and IBM's ACP~\cite{acp}. These protocols provide common interfaces for agent discovery and message exchange, aiming to standardize agentic workflows. 
Figure~\ref{fig:code_listing} shows an example using A2A to connect a ``developer'' agent with a ``testing'' agent. The workflow obtains agent cards (Lines 3 and 7), sends a streaming message to the developer agent (Line 21), and forwards the response to the tester agent (Line 24). 
Infrastructure engineers deploy these workflows and manually tune the communication, batch size, routing scheme, and other parameters to achieve the desired performance.

\subsection{Limitations today}

{\bf Early binding.}
While protocols like A2A simplify agent wiring, they require \emph{early binding} of the communication strategy. Developers must decide at design time whether agents will interact via batching, streaming, or synchronous calls. These decisions (e.g., Lines 22 and 28 in Figure~\ref{fig:code_listing}) are hardcoded into the application, even though the ideal communication mode may vary with runtime conditions. As shown in Figure~\ref{fig:request_rate}, a suboptimal strategy can degrade performance by up to $3.6\times$. 

{\bf Lack of global visibility.}
Existing agentic serving systems provide no mechanism for agentic communication to have visibility into the global serving system state. For example, say there are two instances of an agent, and there is an excessive queue buildup at one instance. Today, it is not possible to inform other upstream agents in the workflow of the existence of such a queue buildup to enable rerouting or changing the communication strategy. %

{\bf Limited end-to-end control.}
As illustrated in Figure~\ref{fig:code_listing}, there is little end-to-end control over runtime communication and serving behavior.  Current abstractions like A2A expose low-level message operations but lack hooks for high-level coordination or optimization across an entire agentic pipeline. For example, it is not possible to express scheduling policies that enable pipeline-wide prioritization of interactive or latency-sensitive requests. Nor is it possible to inject speculative calls or hints~\cite{agarwal2024symphony} (e.g., ahead-of-time movement of KV caches to minimize runtime overhead) to guide workflow execution. 

In essence, the protocols and agentic serving systems today ease the construction of simple pipelines but inhibit adaptation and optimization — especially in dynamic, high-throughput, and low-latency serving environments. %

\subsection{Improving Agentic Serving}

We believe that improving the serving of agentic pipelines requires rethinking the agentic serving architecture from the ground up: with dynamic control over agentic communication based on the current agent- and system-state to optimize operational goals. Realizing this requires the introduction of new agentic communication management building blocks, and new interfaces for %
tools, LLM agents, and serving systems to enable systematic %
integration with %
the building blocks.

\section{A Software-Defined Approach}

\begin{figure}[t!]
    \centering
    \includegraphics[width=0.9\linewidth]{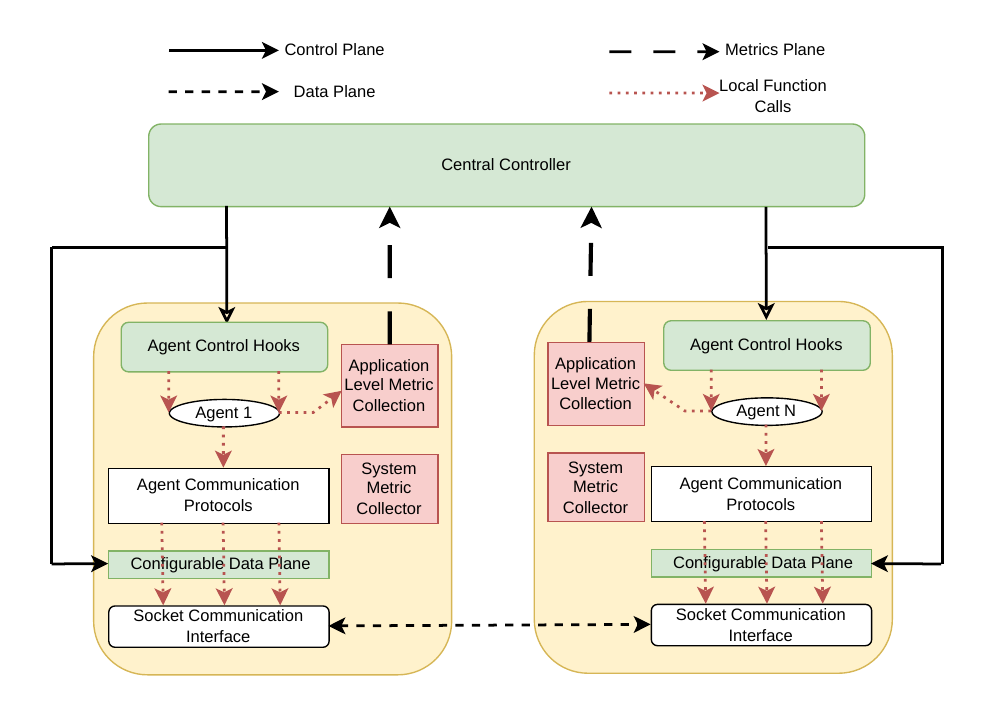}
    \vspace{-10pt}
    \caption{\small{\bf{Our proposal: The control plane orchestrates both data plane and agent/tool actions based on global telemetry.}}}
    \vspace{-10pt}
    \label{fig:agent-control-schematic}
\end{figure}

The above need is likely familiar to the reader -- it is reminiscent of how SDN opened control interfaces to otherwise black-box routers/switches, enabling tighter, network-wide optimization and management. We seek a similar transformation of agentic serving. Specifically, we propose building a software-defined agentic serving stack (Figure~\ref{fig:agent-control-schematic}). 

We propose three components for efficient agentic inference. 
A centralized control plane, which makes data driven decision to control communication and other serving characteristics of each agent. Second, a metrics plane for collecting data and making it available to the controller to make informed decisions. Finally, a data plane substrate that allows for controller to perform fine-grained control.

\subsection{Agent-Serving Control Plane}
\label{sec:control-plane}

Drawing inspiration from SDN, the control plane is responsible for interpreting high-level operator goals and enforcing runtime decisions over agentic serving.  %

{\bf Goals.} We design the control plane to meet three key requirements: (1) \textit{System-wide visibility.} The control plane must have access to all telemetry data exposed by the metrics plane (\S3.2)—including system-level load, application-level performance, and request metadata. This opens the room for fine-grained control and end-to-end optimization.
(2) \textit{Rich control surface.} Beyond managing the data plane (\S3.2), the control plane, realized as a logically central controller, should have hooks into the agent or tool implementations themselves. This enables a broader space of runtime adaptations to improve overall performance and efficiency, such as rescheduling, priority assignment, speculative execution, or cross-node state transfers. (3) \textit{Intent-driven control.} Infrastructure engineers managing the agentic serving platform  should be able to specify high-level goals (e.g., ``maximize throughput under latency bounds’’), and the control plane should compile these into concrete policy rules that evolve the system state toward the goal.

To illustrate the need for a rich control surface with deep agent-level control, consider a scenario where an agent instance becomes overloaded. The controller may choose to reroute a request to another instance. However, if the request relies on key-value (KV) cache state held by the original agent, naive rerouting would require costly recomputation and increase latency. With runtime hooks into the agent, the controller can transfer the relevant state during the handoff—saving compute and improving responsiveness.

A rich control surface is valuable but challenging—it requires integrating with numerous agents and tools, each with unique APIs. This often demands bespoke control plane APIs, making it difficult to build control programs and requiring frequent changes to the controller. Ideally, the control plane should simplify integration through a unified yet effective interface. %

\begin{table}[t!]
\caption{API Interface required by the controller}
\vspace{-10pt}
\label{tab:controller-API}
\resizebox{0.9\linewidth}{!}{
\begin{tabular}{@{}lll@{}}
API                          & Parameters                                 & Description                    \\ \midrule
\multicolumn{1}{l|}{set()}   & \multicolumn{1}{l|}{parameter name, value} & Set  parameter to value        \\
\multicolumn{1}{l|}{reset()} & \multicolumn{1}{l|}{parameter name}        & Set parameter to default value
\end{tabular}}
\vspace{-10pt}
\end{table}

{\bf Design.} We propose a control plane design with the following salient attributes:

   \textit{(1) Data plane control.} Each node runs a configurable shim (as described in \S\ref{sec:data-plane}) that can receive and execute control rules from a controller. Rather than micromanaging individual messages, the controller installs rules. These are of two kinds: (a) \textit{Agent-level rules} govern the default communication mode for an agent pipeline (e.g., batch vs. stream, admit only high-priority requests under load). (b) \textit{Request-level rules} provide fine-grained control over the routing of a request to agent instances (e.g., dynamically switch the agent instance that handles a request; or, when an agent sends a speculative request, block it until resources are free).

    \textit{(2) Agent/tool hooks.} Each agent/tool exposes a registration interface during launch, advertising supported control APIs and parameters (e.g., pause, throttle, transfer state, re-prioritize). These APIs allow the controller to take targeted runtime actions that go beyond the communication substrate. However, as discussed earlier, integrating with each tool/agent's unique API can be tedious. In order to overcome this challenge, we argue for the standardized API described in Table ~\ref{tab:controller-API} which every tool and agent needs to support. Our proposed API includes just two functions, a "set" and "reset". At registration time, each agent exposes the various control and API knobs they have. To utilize these knobs, the controller directly calls the "set" function with parameter name and value. The agent or tool on receiving the "set" is required to update the value and automatically determine the function to call. 
    
    Let's take an example of setting the maximum batch size to 4 for an agent in an agentic pipeline; the parameter for this in the case of vLLM is `max\_num\_seqs'. Here, our controller would call the set function on the relevant agent, "set(`max\_num\_seqs', 4)". The agent is responsible for implementing a shim layer to transform the "set" statement parameter to the relevant internal API and update the batch size. This separation of concerns enables the controller to scale and evolve easily, while agentic workflow developers only need to implement the specific shim layer once.

    \textit{(3) Intent-driven control as state management.} As the controller polls data  (\S\ref{sec:metrics-plane}), it populates a logical state store. Control logic---written by an operator and executed on top of the controller---is provided with the relevant view of state from the store. Akin to intent-driven network management in SDN~\cite{jin2015covisor}, we anticipate control logic to implement algorithms to control dataplane communication such that the {\em observed} state is eventually transformed to a {\em target} state that aligns with the operator's intended objectives. For example, an objective such as "ensure the end-to-end latency of 90\% of interactive requests is within a specified SLO" could translate to the controller polling the latencies of such requests, and, further, the control logic determining to (a) demote background agents to synchronous mode, (b) reallocate GPU slots, and (c) transfer session context for critical requests, until the requests' SLO is met. 

\subsection{Metrics Plane}
\label{sec:metrics-plane}

Effective control requires visibility. To guide runtime communication behavior, the control plane must observe both system-level and application-level dynamics. Unlike traditional networking, where switch-local metrics (e.g., queue depth, flow rates) often suffice, agent serving demands richer telemetry %
that includes system-level data like memory and compute utilization, as well as application-level metrics like query completion times and agent dependencies (conditional agent execution).
Further, these metrics differ depending on the components of the agent pipeline, \eg for LLM-based agent serving we need metrics like GPU utilization, time per token (TPT), and time to first token (TTFT). Meanwhile, for tools like the file system, we need access to file access latency, throughput, and disk space utilization.  To contain complexity, the disparate metric collection requires a flexible yet unified metrics plane.

\textbf{Goals.} The metrics plane should provide: (1) \textit{Low-latency, low-overhead access} to critical signals like GPU/CPU utilization, memory pressure, and queue lengths at each serving node. (2) \textit{Application-aware telemetry}, including per-request latency. (3) \textit{Flexible aggregation}
Across different metrics, agents and tools will generate different volumes of data; for example, time to next token is generated every time a token is generated, while time to first token is measured once per LLM query. The metric collection plane needs to provide a variety of aggregation functions based on the controller's requirements.  (4) \textit{Semantic understanding of the metrics} to enable the controller to interpret and act on them appropriately.
For instance, when optimizing for higher throughput, the controller needs to recognize that high GPU utilization is desirable, while a large amount of context switching is not.

\textbf{Design.} We propose a two-tier architecture consisting of: (1) \textit{Local metric collectors} at each node gather both system-level (hardware, kernel, runtime) and application-level (agent tool usage, response times) telemetry. This data is stored in lightweight, shared-memory structures. And, (2) \textit{Centralized polling} used by the control plane to fetch metrics on demand. This avoids constant metric streaming and allows for coordinated collection across nodes when needed. 
(3) \textit{Flexible Aggregation Functions.} To enable aggregation of various different metrics, the metrics plane will provide the ability to write custom aggregation functions.

This design allows the control plane to reason globally about the serving system---identifying both compute, memory, and communication hotspots; projecting SLO violations; and detecting opportunities for speculative execution\footnote{Controller can detect low load on a particular agent and use it to perform partial prefills or speculative execution, leading to reduced runtime.}.
By separating metric collection from decision-making, we preserve modularity: agent developers can expose domain-specific metrics without rewriting control logic.

To enable semantic understanding of the metrics, our solution is to design a limited specification language that highlights the metric behaviors. For example, an agentic workflow developer who creates a function to collect a metric could write a structured file (in JSON, or YAML) associated with that metric; the controller can read this %
file and use the information provided to understand the metric's significance. 
However, given the vast number of metrics across diverse tools and agents, it can be quite challenging to specify %
all possible metrics exhaustively. A solution is for users to write the significance of metrics in natural language in the signature of the function for collecting the metric; the framework can then potentially use an LLM to transform that description into a metric specification~\cite{zhu2024large}.

\subsection{Configurable Data Plane}
\label{sec:data-plane}

We envision an agentic communication dataplane that decouples communication behavior from static application logic and enables runtime adaptation.

\textbf{Goals.} The data plane substrate must support different message granularities (e.g., token-level streaming, message-level batching),  priorities, and pacing strategies.  Crucially, the dataplane must expose a simple interface for these knobs to be externally controllable by a system-wide controller (\S\ref{sec:control-plane}). For example, under high agent load, the data plane might employ message-level sends to amortize inference cost, whereas under interactive conditions, it might revert to token streaming for lower latency. 

\textbf{Design.} Rather than reinventing agent communication protocols from scratch, we argue for building compatibility into the design. Our proposed data plane acts as a {\em shim layer} between high-level agent protocols (e.g., A2A, MCP, ACP) and the underlying socket or transport layer. This preserves developers’ ability to use familiar and rapidly maturing agentic APIs while exposing a new point of control: a reconfigurable shim that mediates message attributes. The data plane provides a standardized control interface, regardless of the specifics of the upper-level protocol or lower-level transport. %

\subsection{Strawman Implementation}
We implemented a prototype of the proposed design and experimented with a workflow built using Google A2A. %
In this prototype, we modified the A2A protocol to invoke our custom communication library, built on gRPC, which supports synchronous, asynchronous, and streaming communication. Our metrics plane tracks the current load on each agent/tool instance. As before, we use one developer agent and one Testing agent running the MetaGPT workload. 

\begin{figure} [t!]
\begin{minipage}[t]{0.21\textwidth}
    \centering
    \includegraphics[width=0.9\linewidth]{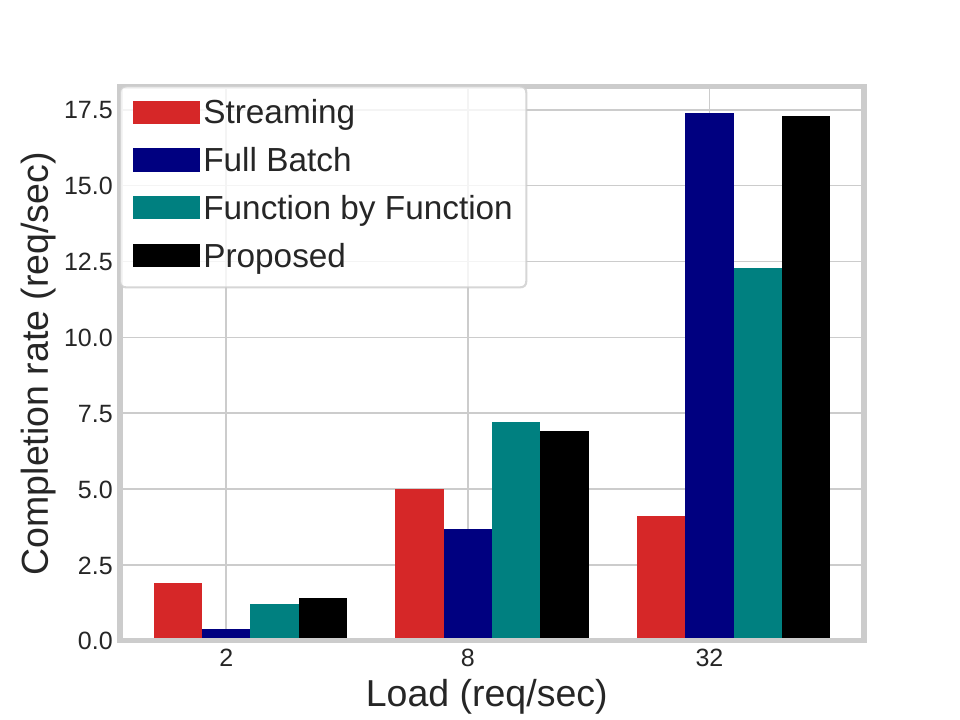}
    \vspace{-10pt}
    \caption{\small{Request rate with control: Using control for message communication granularity. }}
    \vspace{-10pt}
    \label{fig:agent-control-throughput}
\end{minipage}\quad
\begin{minipage}[t]{0.21\textwidth}    
    \centering
    \includegraphics[width=0.9\linewidth]{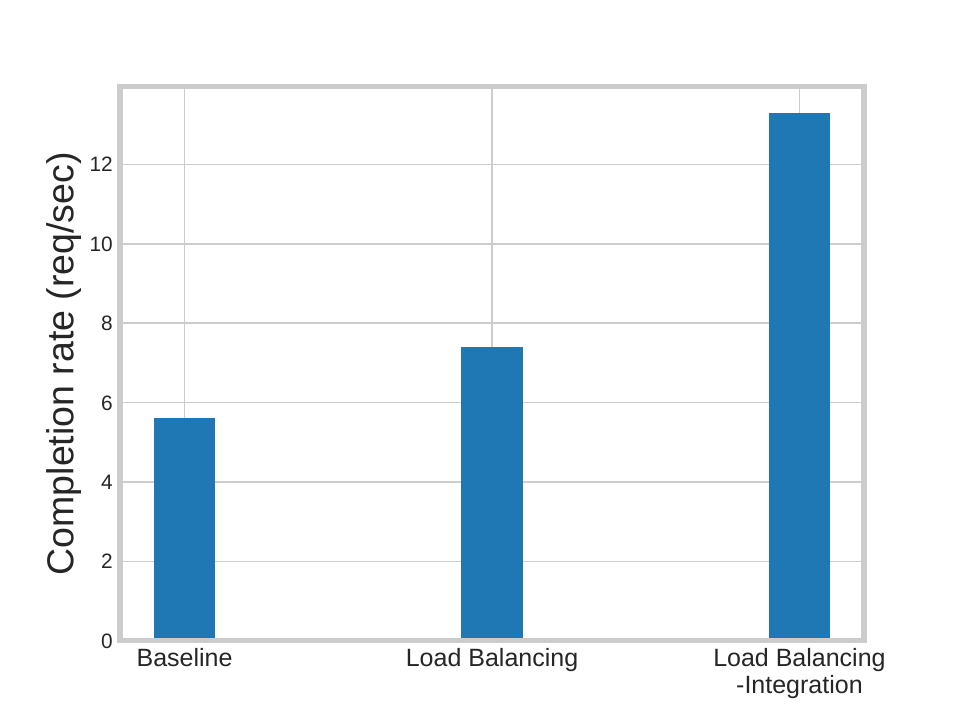}
    \vspace{-10pt}
    \caption{\small{Controller with hint Integration: We use hints to move K,V cache at the time of load balancing.}}
    \label{fig:controller-integration-throughput}
    \end{minipage}
    \vspace{-10pt}
\end{figure}

{\bf Communication Control.}
The controller is designed to adapt the communication pattern dynamically based on load.
As shown in Figure~\ref{fig:agent-control-throughput}, our framework enables switching communication strategy in response to system load to quickly converge on the most effective mechanism.

{\bf Extended control surface.}
To show the impact of giving controller access to the underlying agent serving control. We modify the underlying serving framework (vLLM~\cite{vllm}) to provide a mechanism that allows for the transfer of KV caches from one instance of the agent to another. 
We launch one instance of the software engineering agent and two instances of the testing agent. 
We program the controller to minimize load imbalance across the test agent instances. 
We try two cases, one with integration, where the controller preemptively sends the testing agent a hint that a request will arrive, and the second where the test agent can transfer the KV cache from another instance {\em after} the request arrives.
We observe in Figure~\ref{fig:controller-integration-throughput} that hints perform $1.8\times$ better than without hints. 
We also observe that our proposed solution performs $2.3\times$ better than the baseline with no load balancing.

\section{Related Work}
Agentic LLM systems have received significant attention due to their rapid adoption. A majority of the work has focused on three broad areas: algorithms for prompting and improving the performance of agentic systems; enhancing the efficiency of agent serving; and designing protocols for agent and tool communication.

{\bf Algorithmic improvements.}
Agentic LLM algorithms are still rapidly evolving. This algorithmic evolution can significantly change the compute and control flow in the agentic application pipelines. For example,the dataplane must expose a simple interface for these knobs MetaGPT~\cite{hong2023metagpt}, assigns roles like "developer", "tester", "program manager", etc., to different LLMs to generate code; meanwhile, recent work to improve code generation~\cite{wang2024executable} requires access to a live environment. This algorithmic evolution significantly impacts the workload, as now different agents have to wait for the execution environment before proceeding; depending on the complexity of the code, this can take minutes. Our framework's control plane, through effective resource orchestration, can hide the complexities arising from such algorithmic evolution (e.g., share GPU with multiple concurrent requests). 

{\bf System Improvement.}
Prior work~\cite{luo2025autellix, mei2024aios} has looked at improving the efficiency of agentic systems. However, these systems typically target a specific type of agent workload and custom-developed policies for efficient serving. Our work, rooted in the SDN philosophy, applies generally to a wide range of agent workflows. 

{\bf Agent Communication.}
There has been significant interest in developing agentic communication protocols---ACP~\cite{acp}, A2A~\cite{a2a}, and MCP~\cite{mcp} are notable examples---with %
the goal of standardizing communication across %
diverse agents. In our work, rather than proposing a new protocol, we demonstrate that agent serving necessitates rethinking, with clear unifying interfaces that enable effective control and management toward various system-wide goals, including efficiency.

\section{Discussion and Conclusion}
\label{sec:discussion}

We briefly discuss natural extensions of our proposal first, followed by interesting open questions it brings to fore.

{\bf Supporting External Agents.}
Agentic workflows often span both user-managed infrastructure and external cloud-hosted tools~\cite{}. Our framework extends naturally to such hybrid setups. The controller can factor in latency, cost, or rate limits when invoking external services, choosing configurations that optimize for both performance and budget. Even partial integration—via standardized APIs and light-weight shim layers—can yield measurable benefits.

{\bf Online Policy Adaptation.}
With global visibility into request context and system metrics, the controller can adapt policies at runtime to reduce SLO violations. For instance, if a verifier in the pipeline triggers regeneration, the controller can raise the request’s priority or reroute it for faster execution. This ability to adjust on the fly is key to managing dynamic, multi-agent pipelines.

Our proposal is a starting point for exercising systematic control over agentic serving systems. In addition to discussions around the interfaces and APIs we propose, our work raises new questions across language design, control-agent interfaces, and metrics infrastructure.

{\bf Languages for Agentic Control.} We envision declarative policy languages that let infrastructure engineers express high-level goals—e.g., ``stream under 200ms latency'', ``avoid speculative calls under load'', ``transfer cache if queue depth is high''. Unlike SDN, agentic systems require semantics that span task structure, latency, agent roles, and compute constraints. Key challenges include defining sufficiently expressive primitives, incorporating real-time cross-agent feedback, and reasoning about policy correctness. 

{\bf Control-Agent Interfaces.}
Effective runtime control demands richer interfaces between the controller and agents/tools. A general interface that provides more control than our current proposal would need more capabilities, e.g., interrupting planning loops, cloning context, or suppressing non-critical subcalls.

Agents differ in structure and autonomy, so control interfaces must be both flexible and predictable. One path forward is to define capability classes (e.g., stateless tools vs. stateful planners) with standard control APIs. However, open questions currently remain around capability discovery, policy scope, and conflict resolution.  Nevertheless, we believe that our proposed solution of having standard API's and letting users write a small shim layer to integrate the controller API with agent or tool internal APIs would significantly reduce the effort to introduce new agents. 

{\bf Metrics Infrastructure.}
Our design assumes timely, structured telemetry, but collecting this at scale is nontrivial. First, agent-serving spans heterogeneous backends (vLLM, Triton, custom runtimes), making consistent instrumentation difficult. Second, metrics often require causal tracking across long-lived, multi-agent pipelines. Third, control loops need low-latency access to high-volume telemetry. While our initial prototype advocate aggregation functions to manage volume, new abstractions are needed to expose relevant signals at useful fidelity to the control plane without overwhelming the system or undermining control.

\bibliographystyle{plain}
\bibliography{ref}

\end{document}